\documentclass[aps,twocolumn,floatfix,superscriptaddress]{revtex4-2}

\usepackage{booktabs}
\usepackage{amsmath,amssymb,amsthm,bm}
\usepackage{bbm}
\usepackage{braket}   

\usepackage{graphicx}
\usepackage{float}
\usepackage{tikz}
\usetikzlibrary{decorations.pathreplacing,calc}
\usepackage{quantikz}

\usepackage{multirow}
\usepackage{enumitem}
\usepackage{xcolor}

\usepackage[colorlinks]{hyperref}
\hypersetup{
  bookmarksnumbered,
  pdfstartview={FitH},
  citecolor=blue,
  linkcolor=red,
  urlcolor=black,
  pdfpagemode=UseOutlines
}
\usepackage{cleveref}
\usepackage{xcolor}

\usepackage{xparse}

\usepackage[newcommands]{ragged2e}

\theoremstyle{definition}

\newtheorem{theorem}{Lower Bound}
 \newtheorem*{theorem2*}{Upper Bound}
\allowdisplaybreaks

\makeatletter
\newenvironment{proof*}[1][\proofname]{\par
  \pushQED{\qed}%
  \normalfont \partopsep=\z@skip \topsep=\z@skip
  \trivlist
  \item[\hskip\labelsep\itshape #1\@addpunct{.}]\ignorespaces
}{%
  \popQED\endtrivlist\@endpefalse
}
\makeatother

\ExplSyntaxOn
\NewDocumentCommand{\mref}{m}{\quinn_mref:n {#1}}
\seq_new:N \l_quinn_mref_seq
\cs_new:Npn \quinn_mref:n #1
 {
  \seq_set_split:Nnn \l_quinn_mref_seq { , } { #1 }
  \seq_pop_right:NN \l_quinn_mref_seq \l_tmpa_tl
  ( 
  \seq_map_inline:Nn \l_quinn_mref_seq { \ref{##1},\nobreakspace }
  \exp_args:NV \ref \l_tmpa_tl
  ) 
 }
\ExplSyntaxOff

\begin{document}
\title{Observable measures of multipartite entanglement}
\author{Francois Payn}
\affiliation{DISAT, Politecnico di Torino,  Torino 10129, Italy}
\author{Davide Girolami}
\affiliation{DISAT, Politecnico di Torino,  Torino 10129, Italy}

\date{\today}

\begin{abstract}
Multipartite entanglement is the premier resource for quantum technologies. Yet, its exact quantification in the laboratory is notoriously challenging, typically requiring the full knowledge of high-dimensional quantum states. Here, we construct   observable bounds to  multipartite entanglement for systems of arbitrary size, which are functions of the local and global state purities, and correlation functions.
First, we derive experimentally accessible upper and lower limits to  both the bipartite entanglement of formation and the squashed entanglement of bipartite systems, by leveraging cornerstone results of quantum information theory: the entropy strong subadditivity inequality and the Koashi-Winter monogamy relation. Then, we convert them into bounds to the entanglement up to degree k for arbitrary states, and to the genuine k-partite entanglement, by employing a recently proposed method. Finally, we analytically and numerically test these results, by bounding the multipartite entanglement of several relevant states and mixtures, including the important classes of GHZ, Dicke, W states, and random pure states.
\end{abstract}

\maketitle

\section*{Introduction}

Multipartite entanglement is the defining feature of many-body quantum systems, the building blocks of Nature and the hardware of quantum technologies \cite{Jozsa-Linden,Horodecki_entanglement}. However, its quantification is a difficult theoretical and experimental problem \cite{sep_multi_par}, due to the large number of degrees of freedom in quantum systems   
 \cite{sep_crit_multi, detect_multi, detect_multi_2}. \\Significant  effort has been devoted to studying its strongest form, \textit{genuine multipartite entanglement}, since it can be exploited for various quantum information and computational tasks \cite{gme1,gme2,gme3,gme4}. Yet, dissecting the inner correlation structure of multipartite systems, i.e., multipartite entanglement of degree $k<N$ in $N$-particle states, calls for  sophisticated mathematical and experimental tools that cannot be directly inherited from bipartite entanglement theory.\\ \\
In a recent work \cite{oldpaper}, we introduced an operational definition of multipartite entanglement, which departs from the vastly studied classifications of quantum states in terms of separability and producibility \cite{sep1,sep2,prod,prod2,prod3}.\\ The idea is to construct multipartite entanglement measures from bipartite ones by summing bipartite entanglement across partitions of different clusters. As a result, one is able to quantify multipartite entanglement of any degree $k$ in terms of a parent bipartite entanglement monotone. \\
 The proposal has two main merits. First, its validity is independent of the chosen bipartite entanglement measure: regardless of it, the protocol outputs a measure of $k$-partite entanglement that meets the canonical properties of quantum correlations quantifiers. Second, there is an underlying compelling interpretation of the resulting definition of $k$-partite entanglement: its {\it ex nihilo} creation, i.e., from any possible separable state, requires engineering at least $k-1$ quantum communication channels.\\
Yet, as expected,  $k$-partite entanglement turns out to be analytically computable only in  a few simple cases, with the notable exception of  Greenberger--Horne--Zeilinger (GHZ) states and W states of any dimension \cite{GHZ,W_state,W1,W3}, for which we have been able to find the closed formula of the $k$-partite entanglement of formation \cite{smolin}. Further,  as calculating the exact value of any bipartite entanglement quantifier usually requires full state tomography of experimentally prepared states, estimation in the laboratory of parent multipartite entanglement measures is equally hard.
This limitation calls for investigating computationally and experimentally friendly approximations of $k$-partite entanglement measures, as it has been extensively done for the established notions of multipartite entanglement based on the concept of separability \cite{exp11,exp2,exp3,exp4,exp5,exp6,exp7,exp8,entmeasures6,entmeasures7,entmeasures8,entmeasures9,entmeasures10,entmeasures11}.\\ \\
Here, we calculate observable upper and lower bounds to the recently defined $k$-partite entanglement of formation and the $k$-partite squashed entanglement \cite{squash}. The result holds for arbitrary quantum states of finite-dimensional systems.  
First, we bound bipartite entanglement of formation and squashed entanglement through experimentally accessible quantities requiring low operational cost. We build them by employing a wide range of information-theoretic tools. The first presented lower bound follows from the sharpest form of the strong-subadditivity inequality of the von Neumann entropy \cite{lieb}, and it is a function of local and global state purities. A different lower bound is obtained by leveraging a lower bound to the mutual information in terms of covariances of local observables \cite{hastings}. Finally, we calculate an upper bound to the entanglement of formation by leveraging the Koashi-Winter monogamy relation \cite{koashi_winter}.\\
  Then, we convert these results  into bounds to multipartite entanglement measures, and test their usefulness in several case studies. Notable examples are the GHZ states  and W states. They represent two different classes of entangled states, i.e., they cannot be transformed into each other by Local Operations and Classical Communication (LOCCs)  \cite{sep2}. They have found applications in diverse tasks including quantum metrology \cite{quant_metro}, quantum secret sharing \cite{secret_sharing}, superdense coding \cite{W2}, and quantum teleportation \cite{quant_telep}. This motivates the interest in certifying their experimental implementation. Such characterization may, for instance, help determine which type of state has been produced in a preparation procedure or how much decoherence a given state has undergone. Also, we test our bounds to evaluate entanglement in random pure qubit states.\\
The paper is structured as follows. In Section \ref{sec2}, we derive the purity-based upper and lower limits to the bipartite entanglement of formation and the squashed entanglement. Then, we review the theoretical characterization of multipartite entanglement as sum of bipartite entanglement across bipartitions of different sizes  (Section \ref{sec1}), which we proposed in \cite{oldpaper}. Building on this result, we derive bounds to multipartite entanglement and  explicitly compute them in several cases, as discussed in Section \ref{sec3}. 
Finally, we draw our conclusion and outlook potential follow-ons studies.

\section{Observable bounds to bipartite entanglement}\label{sec2}
Entanglement is a non-linear property of quantum states. Therefore, there is no Hermitian operator whose average value is a fully satisfactory entanglement measure. Further, even when full knowledge of a system density matrix is available, entanglement measures are hard to compute. These well known issues have been addressed by defining entanglement witnesses \cite{exp0,exp1}. Yet, the estimation of quantitative bounds, rather than the bare confirmation that entanglement exist in our devices, requires more sophisticated information-theoretic lower and (possibly) upper limits to known entanglement monotones.\\ 
Here, we consider two important entanglement quantifiers. The entanglement of formation of a state $\rho_{AB}$ is the minimal average pure state entanglement (the local von Neumann entropy $S(\rho_{A(B)}):=-\text{Tr}[\rho_{A(B)}\log \rho_{A(B)}]$) computed over all possible pure state decompositions \cite{smolin}:
  \begin{align}\label{eof}
      E_F(\rho_{AB}):=\min_{p_i,\psi_{i,AB}: \rho_{AB}=\sum_i p_i\ket{\psi}\bra{\psi}_{i,AB}}\sum_i p_i E(\ket{\psi}_{i,AB})
\end{align}
while the squashed entanglement is the minimal conditional mutual information 
\begin{align}
&I(A:B|C):= I(A:BC)-I(A:C),\nonumber\\
&I(A:B):= S(\rho_A)+S(\rho_B)-S(\rho_{AB})
\end{align}
 over all the possible tripartite extensions \cite{squash},
\begin{equation}
    E_{\text{sq}}(\rho_{AB}):= \frac{1}{2} \inf_{\rho_{ABC}: \rho_{AB}=\text{Tr}_{C}\rho_{ABC}} I(A : B \mid C)
\end{equation}
While the two quantities reduce to the local entropy for pure states, both optimizations are generally computationally intractable if the state is mixed. There exists a closed formula for the entanglement of formation for two-qubit mixed states, and a few other special cases \cite{wootters}. There is also  an analytical upper bound for qubit-qudit states \cite{osborne}, and an analytical exact form for  W states and its marginal states \cite{oldpaper}. The squashed entanglement is even harder to calculate: no non-trivial states admit a known closed formula.\\
Notably, one has \cite{squash}
\begin{equation}\label{sq_form_rel}
    E_{sq}(\rho_{AB}) \leq E_F(\rho_{AB})
\end{equation}
Hence, a lower bound to the squashed entanglement is also a lower limit to the entanglement of formation, and an upper bound to the latter is also always larger than the former. \\
Thus, we can calculate the range of possible values of both quantities by means of  lower bounds on the squashed entanglement and upper bounds to the entanglement of formation. A crucial requirement is that these functions should depend on directly measurable quantities. The customary choice is the state purity, as it can be extracted by measuring the SWAP  operator $V$ on two copies of a state: $\text{Tr}[\rho^2]=\text{Tr}[V \rho\otimes \rho]$ \cite{exp1}, or by the more recent method of shadow tomography \cite{huang,science,xiao}. \\
Our first result is
\begin{theorem}\label{low1}
{\it Given a bipartite system $AB$ in a state $\rho_{AB}$, one has
    \begin{align}
    E_{\text{sq}}(\rho_{AB}) &\geq \max\,\bigl\{
    L(\text{Tr}[\rho_{A(B)}^2]) - U\left(\text{Tr}[\rho_{AB}^2],d_{AB}\right),0
    \bigr\} \nonumber
    \end{align}
where $L(\text{Tr}[\rho_X^2])\leq S(\rho_X)\leq U(\text{Tr}[\rho_X^2],d_X)$ are bounds to the Von Neumann entropy of $\rho_X$,  and $d_X$ is the dimension of the system $X$ Hilbert space.}
\end{theorem}

\begin{proof*}
 The sharpest form of the entropy strong subaddivity inequality, due to Carlen and Lieb \cite{lieb}, is given by
 \begin{align}\label{cl}
 I(A : B \mid C) \geq 2 \max\{S(\rho_{A(B)}) - S(\rho_{AB}), 0\}.
\end{align}
Consequently, (half of) the right-hand term is a lower bound to the squashed entanglement, which, crucially,  does not depend on the extension $C$.\\
The Lieb-Carlen bound is saturated for global pure states, and it is generally informative for highly entangled states, when the coherent information $S(\rho_{A(B)}) - S(\rho_{AB})$ becomes positive.\\
Entropies in general are not exactly computable either on paper and in practice, as they demand diagonalization of large matrices.  Yet, their values can be constrained by  computable parameters, e.g.,  purities of states $\rho_{A(B)}$, and $\rho_{AB}$ and the involved system dimensions.\\ Specifically, by employing the method developed in \cite{xiao}, we evaluate upper and lower bounds on the entropy of a single subsystem as the maximal and minimal entropy compatible with a fixed purity $P_X := \mathrm{Tr}[\rho_X^{2}]$:
\begin{align}\label{lower_ent}
    S(\rho_X) \geq \min_{\sum_i{\lambda_i^2} =P_X}{\left(-\sum_i{\lambda_i \log \lambda_i}\right)} &=: L(P_X)\nonumber\\
    S(\rho_X) \leq \max_{\sum_i{\lambda_i^2} = P_X}{\left(-\sum_i{\lambda_i \log \lambda_i}\right)} &=: U(P_X,d_X)
\end{align}
where $\lambda_i$ are the eigenvalues of the state $\rho_X$ in the Hilbert space of dimension $d_X$. A few algebra steps yield the following results
\begin{align}
&L(P_X)=
   -\!\left[
  (1 - \lambda_k)\ln(\lambda_1)
  + \lambda_k \ln(\lambda_k)
  \right]\nonumber \\\\
&U(P_X,d_X)
  = -\!\left[
  (1 - \lambda)\ln\!\left(\frac{1 - \lambda}{d - 1}\right)
  + \lambda \ln(\lambda)
  \right] \nonumber
\end{align}
in which
\begin{align}
\lambda &= \frac{1}{d} 
  + \sqrt{\left(1 - \frac{1}{d}\right)\!\left(P_X - \frac{1}{d}\right)}\nonumber\\
\lambda_k &= \frac{1}{k} 
  - \sqrt{\left(1 - \frac{1}{k}\right)\!\left(P_X - \frac{1}{k}\right)} \nonumber\\
\lambda_1 &= \frac{1 - \lambda_k}{k - 1}, \qquad k\in \mathbb{N}: 1/k \leq P_X \leq \frac{1}{k-1}.
\end{align}
Then, the squashed entanglement is always greater or equal than  
$\max\{L(P_{A}),L(P_B)\}-U(P_{AB},d_{AB}))$, and the claim is proven.
\end{proof*}
This result provides an estimate to the entropy in terms of the simplest polynomial function of a quantum state, i.e., the state purity. It is well known that the entropy can be approximated with arbitrary precision by writing down its Taylor expansion in terms of higher-order polynomials $\text{Tr}[\rho^k], k>2$, but those quantities requires an exponentially increasing number of measurements and multiple state copies to be extracted from experimental data \cite{daley}.\\
Second, under operationally motivated constrained, e.g., the global state is almost pure and highly entangled, these bounds become sharp.\\ \\
The rich structure of quantum systems dictates to explore multiple, different bounds to entanglement.  Here, we define an alternative, complementary observable lower limit to the squashed entanglement:
\begin{theorem}\label{low2}
{\it For any bipartite state $\rho_{AB}$ one  has
    \begin{align}
    E_{\text{sq}}(\rho_{AB}) &\geq \max\Bigl(\frac{C(M_A,M_B)^2}{4\|M_A\|^2\|M_B\|^2} - \frac{U(P_{AB},d_{AB})}{2}, 0\Bigl) ,\nonumber 
\end{align}
  in which the correlation function  $C(M_A,M_B) = \langle M_A \otimes M_B\rangle - \langle M_A\rangle \langle M_B \rangle$, $\langle X\rangle:=\text{Tr}[\rho_{AB} X],$ captures  the (linear) statistical dependence between two local observables $M_A$ and $M_B$, and $||X||$ is the operator norm of $X$.}
\end{theorem}

\begin{proof*}
The Carlen-Lieb inequality in Eq.~(\ref{cl}) implies
\begin{align} \label{sdoppiata}
    I(A:B \mid C) &\geq S(\rho_A) + S(\rho_B) - 2S(\rho_{AB}) \nonumber \\
    &= I(A:B) - S(\rho_{AB}).
\end{align}
We now employ  the result proven in \cite{hastings}: for any two local Hermitian operators $M_A,M_B$, one has
\[
 I(A:B) \geq \frac{C(M_A, M_B)^2}{2\|M_A\|^2 \|M_B\|^2}
\]
Since we previously demonstrated that $ -S(\rho_{AB}) \geq -U(P_{AB},d_{AB})$, the statement is proven.
\end{proof*}
We now report a third result, which provides a universal, analytical upper bound on the entanglement of formation and, consequently, to the squashed entanglement:
\begin{theorem2*}\label{up}
{\it Given a bipartite state $\rho_{AB}$ the entanglement of formation between subsystems $A$ and $B$ satisfies
\begin{align}\label{up}
E_F(\rho_{AB})\leq \sum_i p_i\, S(\rho_{A,i}),
\end{align}
where $\rho_{A,i}=\text{Tr}_C(\mathbb{I}_A \otimes \Pi_{C,i}\rho_{AC}\mathbb{I}_A \otimes \Pi_{C,i})$ is the post-measurement state of $A$ after an arbitrary projective measurement  on a subsystem $C$, such that $\rho_{AB} = Tr_C[\rho_{ABC}]$.
}
\end{theorem2*}

\begin{proof*}
Let us recall the Koashi--Winter relation \cite{koashi_winter},
\begin{equation}
E_F(\rho_{AB}) \leq S(\rho_A) - J(A:C),
\end{equation}
where $J(A\!:\!C)$ quantifies the classical correlations between $A$ and $C$  as the maximal mutual information that is left in their bipartite state after a local measurement on $C$ \cite{henderson}:
\begin{align}
&J(A\!:\!C)=\max_{\{\Pi_i\}}\left\{S(\rho_A)-\sum_i p_i\, S(\rho_{A,i})\right\}\\
&p_i = \mathrm{Tr}\!\left[(\mathbb{I}_A \otimes \Pi_{C,i})\rho_{AC}\right],
\,\rho_{A,i}=\frac{\mathrm{Tr}_C\!\left[(\mathbb{I}_A \otimes\Pi_{C,i})\rho_{AC}\right]}{p_i}\nonumber.
\end{align}
Hence, calling $I(A:C_{\Pi})$ the mutual information between $A$ and $C$ after an arbitrary local measurement on $C$, one has
\begin{align}
J(A:C) &\geq I(A:C_{\Pi})\nonumber\\
&=S(\rho_A)-\sum_i p_i S(\rho_{A,i}).
\end{align}
We then obtain the upper bound on the entanglement of formation in Eq.~(\ref{up}), expressed in terms of local, post-measurement entropies. 
 
\end{proof*}
We observe that this bound requires estimating the local entropy of the post-measurement state $\rho_A$, which is not directly computable, as $A$ can be arbitrarily large, in principle.   However, in Sections \ref{sec1},\ref{sec3} we will show that multipartite entanglement in $N$-particle systems can be quantified by means of bounds to bipartite entanglement between a single component, say $A$, and a subset of the other particles. In particular, in an $N$-qubit system, $A$ will be a single qubit, which makes the upper bound in Eq.~(\ref{up}) easy to compute.

\section{Exact Quantification of Multipartite Entanglement}\label{sec1}

We briefly recall the construction of multipartite entanglement measures in \cite{oldpaper}. We will denote clusters of $l$ particles as $\mathcal{X}_l$ and the $l$-th particle of the system as $\mathcal{X}_{[l]}$.
Consider an $N$-particle state $\rho_N$ and fix $k\le N$.
We select a cluster
$\mathcal{X}^1_k=\{\mathcal{X}^1_{[1]},\ldots,\mathcal{X}^1_{[k]}\}$
of $k$ particles and evaluate the bipartite entanglement between the first particle and the rest of the cluster,
\[
E[\rho_N]\big(\mathcal{X}^1_{[1]}:\mathcal{X}^1_{[2]}\cdots
\mathcal{X}^1_{[k]}\big),
\]
where $E$ is an arbitrary bipartite entanglement quantifier. We then choose another $k$-particle cluster $\mathcal{X}^2_k$ that does not contain $\mathcal{X}^1_{[1]}$
(generally $\mathcal{X}^1_{[i]}\neq\mathcal{X}^2_{[i]}$), and evaluate the analogous bipartite contribution. Iterating this procedure generates a sequence of bipartitions involving clusters of size $k$ until fewer than $k$ particles
remain. One then considers progressively smaller clusters, down to the final bipartition of two particles.\\
For a given ordering of the particles, this procedure defines a sequence
\begin{align}\label{sequences}
p =\Big\{\, & \big(\mathcal{X}^1_{[1]}:\mathcal{X}^1_{[2]}...\mathcal{X}^1_{[k]}\big),\big(\mathcal{X}^2_{[1]}:\mathcal{X}^2_{[2]}...\mathcal{X}^2_{[k]}\big), \nonumber \\& ...,\big(\mathcal{X}^{N-k+2}_{[1]}:\mathcal{X}^{N-k+2}_{[2]}...\mathcal{X}^{N-k+2}_{[k-1]}\big),\nonumber \\&...,\big(\mathcal{X}^{N-1}_{[1]}:\mathcal{X}^{N-1}_{[2]}\big)
\Big\} 
\end{align}
whose associated sum of bipartite contributions is denoted as
\begin{align}\label{sum_sequence}
s_{N,k}(\rho_N)&:=\Big\{\,  E[\rho_N]\big(\mathcal{X}^1_{[1]}:\mathcal{X}^1_{[2]}...\mathcal{X}^1_{[k]}\big)\nonumber\\
&+E[\rho_N]\big(\mathcal{X}^2_{[1]}:\mathcal{X}^2_{[2]}...\mathcal{X}^2_{[k]}\big)+ \nonumber \\& ...+E[\rho_N]\big(\mathcal{X}^{N-k+2}_{[1]}:\mathcal{X}^{N-k+2}_{[2]}...\mathcal{X}^{N-k+2}_{[k-1]}\big)+\nonumber \\&...+E[\rho_N]\big(\mathcal{X}^{N-1}_{[1]}:\mathcal{X}^{N-1}_{[2]}\big)
\Big\}. 
\end{align}
Let ${\cal S}_{N,k}$ be the set of all these
sums obtained from all possible valid sequences of particles.
The total amount of entanglement up to degree $k$ is then defined as
\begin{align}
E^{2\leftrightarrow k}(\rho_N)
:=\max_{s_{N,k}\in{\cal S}_{N,k}} s_{N,k}(\rho_N).
\label{eq1}
\end{align}
This quantity meets the canonical properties of entanglement measures:
\begin{itemize}
\item Faithfulness: If and only if a quantum state  is  a mixture of partially separable states
\begin{align}  
\sum_i p_i\rho_{{k_1,i}}\otimes\rho_{{k_2,i}}\otimes
\ldots\otimes\rho_{k_\alpha,i}, \, k_{1,2,\ldots,\alpha}\leq k,
\end{align}
 then $E^{l>k}(\rho_N)=0, E^{2\leftrightarrow N}(\rho_N)=E^{2\leftrightarrow k}(\rho_N)$;
 \item Invariance under single particle unitary;
 \item Monotonicity under LOCCs on any pair of particles.
 \end{itemize}
By construction, we define genuine $k$-partite entanglement by
\begin{equation}\label{eq2}
E^{k}(\rho_N)
:=E^{2\leftrightarrow k}(\rho_N)
 -E^{2\leftrightarrow k-1}(\rho_N).
\end{equation}
Computing this quantity for a generic state may be an intractable problem, but, when the system is invariant under particle exchange, it  takes the compact form:
 \begin{align}\label{compact}
 E^{k}(\rho_N)=&(N-k+1)\{E[\rho_N](\mathcal{X}_{[1]}:\mathcal{X}_{[2]}\ldots\mathcal{X}_{[k]})\\\nonumber
 -& E[\rho_N](\mathcal{X}_{[1]}:\mathcal{X}_{[2]}\ldots\mathcal{X}_{[k-1]})\},
\end{align} 
where $(\mathcal{X}_{[1]}:\mathcal{X}_{[2]}\ldots\mathcal{X}_{[k]}) \,$and$ \,(\mathcal{X}_{[1]}:\mathcal{X}_{[2]}\ldots\mathcal{X}_{[k-1]})$ are arbitrary sequences.\\
While we cannot expect to capture all the intricacies of multipartite entangled states by  a unique scalar quantity, the ``Entanglement vector'' $\Big\{E^2(\rho_N),E^3(\rho_N),...,E^N(\rho_N)\Big\}
$ provides a full-fledge description of their correlation structures (see Table 1 of \cite{oldpaper} for examples).

\section{Observable Bounds to Multipartite Entanglement}\label{sec3}
\subsection{General framework: bounds to multipartite entanglement from limits to bipartite entanglement}
The protocol described in the previous Section enables one to define  multipartite entanglement measures from an arbitrary bipartite entanglement quantifier, including the entanglement of formation \cite{smolin} 
 and the squashed entanglement \cite{squash}. 
Hence, by employing the observable bounds to these quantities, as derived in Section \ref{sec2}, we can compute observable upper and lower limits on the {\it multipartite entanglement of formation} and the {\it multipartite squashed entanglement}. They are functions of bounds on  bipartite entanglement computed across any bipartition of the system under scrutiny:\\

\noindent {\bf Main Result.} {\it We construct bounds to the total entanglement up to degree k in Eq.~(\ref{eq1}), employing the entanglement of formation or the squashed entanglement as bipartite measure. Instead of optimizing the sum (\ref{sum_sequence}), we optimize separately the sum of bipartite \textbf{Upper Bounds} as given by Eq.~(\ref{up}) and lower bounds given by the largest between \textbf{Lower Bound 1} Eq.~(\ref{low1}) and \textbf{Lower Bound 2} Eq.~(\ref{low2}).
We call $U^{2\leftrightarrow k}(\rho_N), L^{2\leftrightarrow k}(\rho_N)$ their maximal  values. One therefore obtains lower and upper limits to the multipartite entanglement of formation and the multipartite squashed entanglement:
 \begin{align}\label{main}
 &L^{2\leftrightarrow k}(\rho_N)\leq E_{sq,F}^{2\leftrightarrow k}(\rho_N) \leq U^{2\leftrightarrow k}(\rho_N),\nonumber\\
 &L^{k}(\rho_N)\leq E_{sq,F}^{k}(\rho_N) \leq U^{k}(\rho_N)
 \end{align}
Where, $L(U)^{k}(\rho_N):=L(U)^{2\leftrightarrow k}(\rho_N)-U(L)^{2\leftrightarrow k-1}(\rho_N)$}\\ \\
As mentioned before, the computation of the upper bound is still manageable, regardless of the size $N$ of the global system.
Indeed, if we denote by $\rho_{ABC}$ the full state of the system, $\rho_{AB}$ corresponds to the reduced state of a specific cluster. Thus, subsystem $C$ can be taken to be the remaining qubits that have been traced out, and, within a given cluster, $A$ denotes a single qubit, being $B$ the remaining qubits of the cluster.
 When the cluster coincides with the full system, no physical subsystem $C$ is available. In this case, if the global state is mixed, one may introduce a purifying system to play the role of $C$.
If the global state is pure, the bipartite entanglement is simply given by the reduced entropy of one subsystem, and no bound is required.\\ \\

\subsection{Case studies}
\subsubsection{Dephased GHZ state}
\begin{figure}[H]
    \centering
    \includegraphics[width=0.8\linewidth]{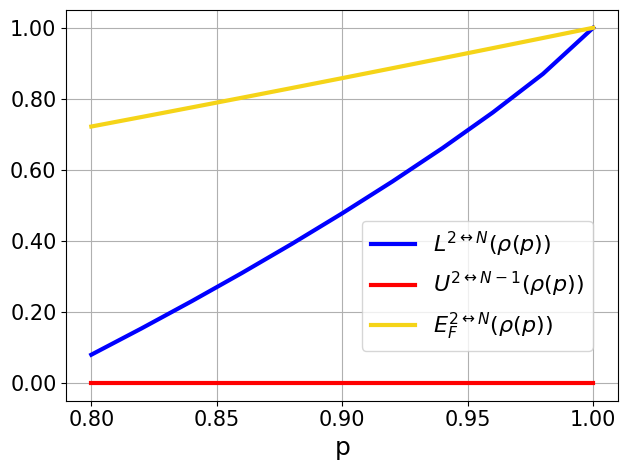}
    \caption{The yellow line represents the  total entanglement of formation $E_F^{2\leftrightarrow N}(\rho(p))$  in Eq.~(\ref{anent}). The blue line shows that the lower bound of the total entanglement of formation increases as $p$ varies from 0.8 to 1, where the state becomes an ideal $N$-partite GHZ state. 
    The red line is the upper bound of $E_F^{2\leftrightarrow N-1}(\rho(p))$, and it is always zero, matching the analytical value.}
    \label{fig1}
\end{figure}

We compute the bounds in Eq.~(\ref{main}) in several interesting cases.\\
Consider the experimental preparation of an $N$ qubit GHZ state affected by depolarizing noise:
\begin{align}\label{noisy_ghz}
&\rho(p)=p\ket{\mathrm{GHZ_N}}\bra{\mathrm{GHZ_N}}+(1-p)\rho^{\mathrm{deph}}_N,
\,\,\, 0\le p\le1, \nonumber\\
&\ket{\mathrm{GHZ}_N}
=\frac{1}{\sqrt{2}}\left(\ket{0}^{\otimes N}+\ket{1}^{\otimes N}\right),\nonumber\\
&\rho^{\mathrm{deph}}_N
=\tfrac12\ket{0}\bra{0}^{\otimes N}
+\tfrac12\ket{1}\bra{1}^{\otimes N}.
\end{align}
It is well known that the GHZ state possesses only $N$-partite entanglement, $E^{k<N}(\rho(p))=0$. That is, if any of the $N$ qubits is lost or traced out, what remains is a separable state.   Since the dephased state is still symmetric under particle exchange and has rank 2 in any partition, we can analytically compute the $N$-partite entanglement of formation $E^N(\rho(p))$ by employing the method reported in \cite{oldpaper}.  
Based on the framework in Section \ref{sec1}, the $N$-th degree entanglement is given by the bipartite entanglement between one qubit $A$ and the remaining $N-1$ qubits $B$ in the state:
\begin{align}
\rho(p)&=\frac12\Big(\ket{0_A}\ket{0_B}\!\bra{0_A}\bra{0_B}+\ket{1_A}\ket{1_B}\!\bra{1_A}\bra{1_B}
\Big)\nonumber\\&\quad+\frac{p}{2}\Big(\ket{0_A}\ket{0_B}\!\bra{1_A}\bra{1_B} +
\ket{1_A}\ket{1_B}\!\bra{0_A}\bra{0_B}\Big)\nonumber
\end{align}
Where,
\begin{align}
&\ket{0}_A =\ket{0},
\qquad \qquad
\ket{1}_A=\ket{1} \nonumber \\
&\ket{0}_B=\ket{0}^{\otimes(N-1)},
\quad \,\,
\ket{1}_B=\ket{1}^{\otimes(N-1)}. \nonumber
\end{align}
The support of subsystem $B$ is two–dimensional. We therefore introduce
the local isometry $V:\mathbb{C}^2\rightarrow(\mathbb{C}^2)^{\otimes(N-1)}$ such that:
\begin{equation}
V\ket{0}=\ket{0}^{\otimes(N-1)},\quad
V\ket{1}=\ket{1}^{\otimes(N-1)}, 
\end{equation}
which preserves bipartite entanglement between A and B. We use such map to reduce the $N$-qubit state to a two-qubit state with the same  entanglement:
\begin{align}
\rho(p)_{2}&=(\mathbb{I}\otimes V^\dagger)\,\rho(p)\,  (\mathbb{I}\otimes V)=
\frac{1}{2}
\begin{pmatrix}
1 & 0 & 0 & p\\
0 & 0 & 0 & 0\\
0 & 0 & 0 & 0\\
p & 0 & 0 & 1
\end{pmatrix}
\end{align}
The entanglement of formation of this state can be computed analytically  \cite{wootters,eberly}:
\begin{align}\label{anent}
&E_F(\rho(p)_{2})=h\!\left(\frac{1+\sqrt{1-p^2}}{2}\right),\nonumber\\
&h(x)=-x\log_2 x-(1-x)\log_2(1-x).
\end{align}
Hence, this is equal to the bipartite entanglement of formation in the dephased GHZ state  $\rho(p)$ across any bipartition, including the 
$1$ qubit vs $N-1$ qubits one.\\
The dephased GHZ state $\rho(p)$
is symmetric under particle exchange. Consequently, we can employ the compact form  in Eq.~(\ref{compact}) to quantify $k$-partite entanglement.\\
Then, we construct the lower bounds by estimating purities and correlation functions of Pauli measurements.  \\
The upper bound  is  obtained, in this case, by 
measuring subsystem $C$ in, say, the computational ($Z$) basis,
producing outcomes $i=\{0,1\}$ with probabilities
$
p_i=\mathrm{Tr}\!\left[(\mathbb{I}_A\otimes\ket{i}\bra{i}_C)\rho_{AC}\right].
$
Conditioned on each outcome, single-qubit tomography is performed on $A$.\\
In Figure \ref{fig1}, we plot the total entanglement of formation $E_F^{2\leftrightarrow N}(\rho(p))$, which we computed analytically, its  lower bound $L^{2\leftrightarrow N}(\rho(p))$,  and the upper bound to the entanglement of degree up to $N-1$, $U^{2\leftrightarrow N-1}(\rho(p))$, for the experimentally motivated values of $p$, ranging from $0.8$ to $1$. \\
The picture highlights how the bounds certify that  there is no entanglement of degree lower than $N$ but an increasing $N$-partite entanglement. Therefore, we  have a genuine yet decohered GHZ state.


\subsubsection{Genuine Multipartite Entanglement Under Particle Symmetry}
\begin{figure}
    \centering
    \includegraphics[width=0.8\linewidth]{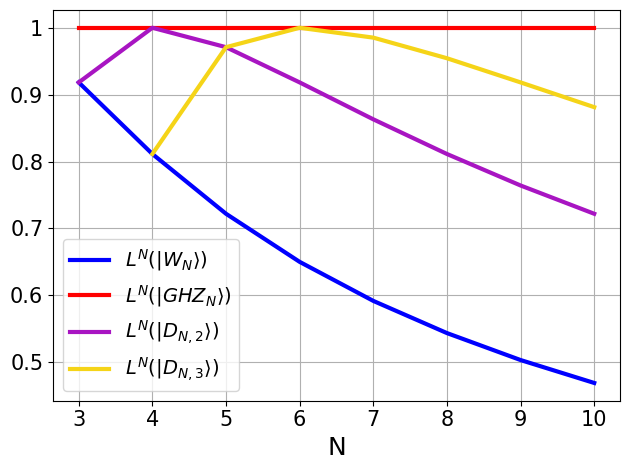}
    \caption{We study the scaling with the particle number $N$ of the lower bound to the genuine $N$-partite entanglement of formation in symmetric pure states. While, as expected, the GHZ state has always one bit of $N$-partite entanglement, the Dicke states display peak entanglement when $N$ is twice the number of excitations. We conjecture this behaviour is universal, i.e., it may hold for any $N$.}
    \label{fig:scaling}
\end{figure}

We now generalize the study to other states that are symmetric under particle exchange. In particular, we track the lower bound to the genuine $N$-partite entanglement of formation, which for a generic $\rho$ are
\begin{equation}\label{genuine_lb}
    L^N(\rho) = L^{2 \leftrightarrow N}(\rho)-U^{2 \leftrightarrow N-1}(\rho),
\end{equation}
in important classes of highly entangled pure states of $N$ particles. Namely, we consider the GHZ state   $\ket{GHZ_N}$, the single excitation W states  $\ket{W_N}$, and the Dicke states $\ket{D_{N,2}}$ and $\ket{D_{N,3}}$:
\begin{equation}
    \ket{W_N} = \frac{1}{\sqrt{N}} \sum_{k=1}^{N} \ket{0}^{\otimes (k-1)} \otimes \ket{1} \otimes \ket{0}^{\otimes (N-k)}
\end{equation}
\begin{equation}
\ket{D_{N,m}} = \frac{1}{\sqrt{\binom{N}{m}}}
\sum_{\substack{x \in \{0,1\}^{N}, \\ \sum_{i=1}^{N} x_i = m}}
\ket{x_1 x_2 \cdots x_N}
\end{equation}
where in general $\ket{D_{N,m}}$ are the Dicke states of $N$ qubits with $m$ excitations.  The choice is dictated by the well-known importance of these states for quantum computing and quantum metrology \cite{W_state,W1,W2,W3,dicke1,dicke2}.\\
We plot in  Fig.~\ref{fig:scaling} the results for systems  from three to ten qubits. An interesting feature of Dicke states is that they can reach the same amount of $N$-partite entanglement of the GHZ state when $N=2\,m$, and then it decreases monotonically by increasing $N$.  Of course, their correlation structure, characterized by $E^k\neq 0, \forall k$, is  hard to study, while, remarkably, the multipartite entanglement of formation is analytically computable for the simpler $W_N$ state, as shown in  \cite{oldpaper}.

\subsubsection{Random Pure States}
We now assume absolute ignorance about the prepared quantum state. This is a rare occurrence, as we usually do know something about the system we want to engineer. Yet, this exercise is instructive,  showing how bounds to the $k$-partite entanglement of formation can be still computable and informative, even when, for example, unknown (e.g., stochastic) noise significantly affects control of quantum particles.\\
Since we do not have any information about symmetries in the system, we are forced to perform the cumbersome maximization detailed in Section \ref{sec1}, in order to calculate
\begin{equation}
    L^k(\rho_N) = L^{2 \leftrightarrow k}(\rho_N)-U^{2 \leftrightarrow k-1}(\rho_N).
\end{equation}
To compute these two terms means, apparently, that one has to check every possible allowed particle sequence (\ref{sequences}). Yet, we  here present a method that computes such values efficiently,  by exploiting the recursive
structure of the construction in Section \ref{sec1}.\\
 At any stage of the protocol, the remaining particles
form a subset \(R \subseteq \{1,\dots,N\}\) with cardinality $|R|=m$. A single step of the sequence corresponds to selecting
\begin{enumerate}
\item a pivot particle $a\in R$,
\item a subset $u_k\subseteq R\setminus\{a\}$ with \(|u_k|=\min(k-1,m-1) \), \label{rule2}
\end{enumerate}
which generates a bipartite contribution associated with the cluster $\{a\}\cup u_k$. Here, we add a lower index $k$ to $u_k$ to keep track of the rule 2. After this step the particle $a$ is removed, leaving one with the reduced set \(R\setminus\{a\}.\) Hence, the optimal summed contribution starting from a generic subset $R$ depends only on the optimal value obtainable from smaller subsets. Defining
\begin{equation}
F_k(R) = \max_{p\,\text{starting from }R} \sum_i E(a^i:u_k^i)
\end{equation}
the maximization is performed by recursion:
\begin{equation}\label{recursion_def}
F_k(R)=\max_{a\in R}\max_{\substack{u_k\subseteq R\setminus\{a\}\\|u_k|=\min(k-1,|R|-1)}}\Big[E_R(a:u_k)+
F(R\setminus\{a\})\Big]
\end{equation}

with boundary condition $F(R)=0$ for $|R|\le1$. Here $E_R(a,u_k)$ denotes the bipartite contribution
evaluated for the cluster generated at that step. This relation replaces the enumeration of complete
sequences by a maximization over subsets. Each subset $R$ is evaluated only once and its optimal
value is stored and reused whenever the same subproblem appears again. Operationally, this corresponds to a ``dynamic programming'' strategy.\\
Our study requires two independent optimizations, in order to calculate lower and upper bounds. Calling $L_{2,R}(a:u_k)$ and $U_{2,R}(a:u_k)$ the lower and upper bound bipartite contributions, we therefore compute two functions,
\begin{align}
F_{\mathrm{L},k}(R) &= \max_{a,u_k} \Big[ L_{2,R}(a:u_k)+ F_{\mathrm{L},k}(R\setminus\{a\}) \Big], \nonumber \\
F_{\mathrm{U},k}(R) &= \max_{a,u_k} \Big[ U_{2,R}(a:u_k) + F_{\mathrm{U},k}(R\setminus\{a\}) \Big],
\end{align}
Where the $\max_{a,u_k}$ is intended as in Eq.~(\ref{recursion_def}). 
The desired quantities are finally obtained as
\begin{align}
&L^{2\leftrightarrow k}(\rho_N)=F_{\mathrm{L},k}(\{1,\dots,N\}), \nonumber \\
&U^{2\leftrightarrow k}(\rho_N)=F_{\mathrm{U},k}(\{1,\dots,N\}).
\end{align}
This procedure   avoids the explicit construction of all particle sequences, reducing the computational cost from sequence enumeration to a recursion over particle subsets. \\

\begin{figure}[h]
    \centering
    \vspace{0.5cm}
    \includegraphics[width=0.8\linewidth]{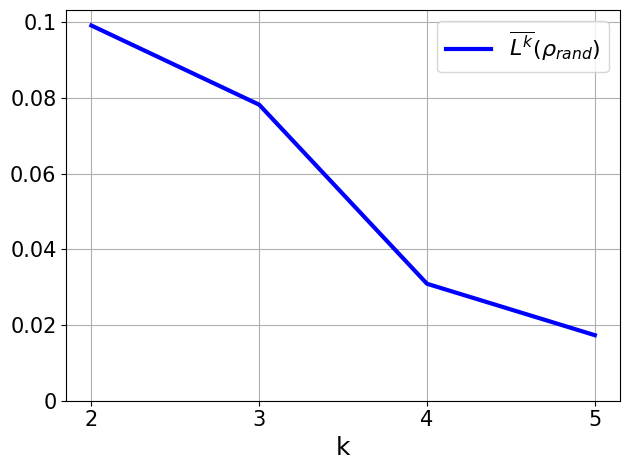}
    \caption{We investigate the   entanglement structure  of random five-qubit pure states, as constructed in (\ref{rand_ket}). 
    In this case, the parameter of the Bernoulli distribution (\ref{bernoulli}) is set to $p=0.85$. 
    For this choice, we calculate the (averaged) lower bounds to the $k$-partite entanglement of formation and squashed entanglement, for $k=2,3,4,5$.}
    \label{fig:rand_phase}
\end{figure}

 We employ this method to explore the entanglement structure of   random pure states. We generate a random sparse pure state in the computational basis $\{ \ket{i} \}_{i=0}^{d-1}$  of an
$N$-qubit Hilbert space $\mathcal{H} = (\mathbb{C}^2)^{\otimes N}$ of dimension $d = 2^N$. For each basis index $i$, we independently draw a Bernoulli random variable
\begin{equation}\label{bernoulli}
    m_i \sim \mathrm{Bernoulli}(1-p)
\end{equation}
which determines whether the corresponding basis component is kept ($m_i=1$) or discarded ($m_i=0$).
If all variables vanish, one index is selected uniformly at random and set to $m_i=1$ in order to avoid the null vector.
For every retained component ($m_i=1$), we assign an independent real coefficient $
a_i \sim \mathrm{Uniform}[0,1]$,
while discarded components carry zero amplitude.
The resulting normalized state reads
\begin{align}\label{rand_ket}
    \ket{\psi_{\mathrm{rand}}}&=\frac{\ket{\tilde{\psi}}}{\sqrt{\sum_{i=0}^{d-1} m_i a_i^{2}}},\\
  \ket{\tilde{\psi}} &=\sum_{i=0}^{d-1} m_i a_i\ket{i} \nonumber.
\end{align}
Defining
$\rho_{rand} = \ket{\psi_{rand}}\bra{\psi_{rand}}$,  we calculate the average lower bound to the $k$-partite entanglement of formation
\begin{equation}
    \overline{L^k}(\rho_{rand}) = \frac{1}{n}\sum_{i=1}^n{L^k(\rho_{rand\,,\,i})},
\end{equation}
where $\rho_{rand\,,\,i}$ denotes one of the $n$ randomly generated states.  We are therefore interested in the average lower bound to the $k$-partite entanglement for random states constructed as $\rho_{rand}$. \\
In Fig.~\ref{fig:rand_phase}, we present results for systems of $N=5$ qubits, averaged over $n=100$ states.

\section*{Conclusion}
In this work, we introduced experimentally accessible bounds to  multipartite entanglement. First, we derived new lower and upper bounds to the bipartite entanglement of formation and  the squashed entanglement, two of the most important entanglement monotones. Then, building on the characterization of multipartite entanglement in terms of bipartite contributions across partitions of different sizes, we computed observables bounds to the ``total entanglement up  to degree $k$'' of an $N$-particle state, and the genuine $k$-partite entanglement.
Notably, the proposed bounds are functions of state purities. That is, their estimation does not require full state tomography, or an arbitrary number of copies of the system state. \\ \\
We demonstrated the usefulness of  these bounds in several case studies.
They allow for non-tomographic certification of GHZ states in presence of depolarizing noise. Also, they discriminate between GHZ states and states exhibiting lower-degree entanglement. 
Remarkably, the required information can be extracted using low-depth measurement schemes, showing that a limited set of observable quantities already suffices to infer the global entanglement structure when decoherence is not too strong.\\
We further employed our method to investigate the scaling of genuine $N$-partite entanglement in families of symmetric states. 
This analysis reveals how multipartite correlations evolve with system size and confirms that the framework captures known structural properties of paradigmatic states while remaining applicable to increasingly large systems.\\
Finally, we showed how our bounds detect multipartite entanglement even in random, non-symmetric pure states.\\
Hence, our results provide an experimentally viable route to probing multipartite entanglement in complex quantum systems, paving the way towards experimental investigations of large-scale quantum correlations. Also, these result could be extended to estimate most general forms of classical and quantum multipartite correlations \cite{tufa}.



\begin{thebibliography}{99}
\bibitem{Horodecki_entanglement}R. Horodecki, P. Horodecki, M. Horodecki, and K. Horodecki, Quantum entanglement, Reviews of Modern Physics 81, 2 (2009).
\bibitem{Jozsa-Linden} R. Jozsa and N. Linden, On the role of entanglement in quantum-computational speed-up, Proceedings of the Royal Society of London.
Series A: Mathematical, Physical and Engineering Sciences 459.2036, pp. 2011–2032 (2003).

\bibitem{sep_multi_par} W. Dür, J. I. Cirac, R. Tarrach, Separability and distillability of multiparticle quantum systems, Physical Review Letters, pp. 3562-3565 (1999).

\bibitem{sep_crit_multi}O. Guehne and M. Seevinck, Separability criteria for genuine multiparticle entanglement, New J. Phys. 12 053002 (2010).

\bibitem{detect_multi}J. D. Bancal, N. Brunner, N. Gisin, and Y. C. Liang, Detecting Genuine Multipartite Quantum Nonlocality: A Simple Approach and Generalization to Arbitrary Dimensions, Phys. Rev. Lett. 106, 020405 (2011).

\bibitem{detect_multi_2}M. Huber, F. Mintert, A. Gabriel, B. C. Hiesmayr, Detection of high-dimensional genuine multi-partite entanglement of mixed states, Phys. Rev. Lett. 104, 210501 (2010).



\bibitem{gme3}
J. L. Beckey, N. Gigena, P. J. Coles, and M. Cerezo, Computable and Operationally Meaningful Multipartite Entanglement Measures, Phys. Rev. Lett. {\bf 127}, 140501 (2021).
 

\bibitem{gme2}J. D. Bancal, N. Brunner, N. Gisin, and Y. C. Liang, Detecting Genuine Multipartite Quantum Nonlocality: A Simple Approach and Generalization to Arbitrary Dimensions, Phys. Rev. Lett. {\bf 106}, 020405 (2011).

\bibitem{gme1}M. Huber, F. Mintert, A. Gabriel, B. C. Hiesmayr, Detection of high-dimensional genuine multi-partite entanglement of mixed states, Phys. Rev. Lett. {\bf 104}, 210501 (2010).
\bibitem{gme4}
M. Huber and J. I. de Vicente, Structure of Multidimensional Entanglement in Multipartite Systems,
Phys. Rev. Lett. {\bf 110}, 030501 (2013). 
\bibitem{oldpaper}F. Payn, M. Minervini, and D. Girolami, Quantifying the Operational Cost of Multipartite Entanglement, arXiv:2602.04760 (2026).


\bibitem{sep2}W. Dür, G. Vidal, and J. I. Cirac, Three qubits can be entangled in two
inequivalent ways, Phys. Rev. A 62, 062314 (2000).

\bibitem{sep1} W. Dür, J. I. Cirac, R. Tarrach, Separability and distillability of multiparticle quantum systems, Phys. Rev. Lett.  83, 3562-3565 (1999).

\bibitem{prod}S. Szalay, k-stretchability of entanglement, and the duality of k-separability and k-producibility, Quantum 3, 204
(2019).
\bibitem{prod2}O. G¨uhne, G. T´oth, and H. J. Briegel, Multipartite entanglement in spin chains, New J. Phys. 7, 229 (2005).
\bibitem{prod3} S. Woelk and O. Guehne, Characterizing the width of entanglement, New Journal of Physics 18, 123024 (2016). 





\bibitem{GHZ}D. M. Greenberger, M. A. Horne, and A. Zeilinger, Multiparticle
Interferometry and the Superposition Principle, Physics Today 46.8, pp. 22–29 (1993).

\bibitem{W_state}W. Dür, G. Vidal, and J. I. Cirac, Three qubits can be entangled in two
inequivalent ways, Phys. Rev. A 62 (2000).
\bibitem{W1}G. Tóth, Multipartite entanglement and highprecision metrology. Phys. Rev. A 85, 022322, (2012). 

 \bibitem{W3}T. Tashima, T. Wakatsuki, S¸. K. Ozdemir, T. Yamamoto, ¨
M. Koashi, and N. Imoto, Local transformation of two
einstein-podolsky-rosen photon pairs into a three-photon
W state, Phys. Rev. Lett. 102, 130502 (2009).


\bibitem{smolin}C. H. Bennett, D. P. DiVincenzo, John A. Smolin, and William K. Wootters, Mixed-state entanglement and quantum error correction, Physical Review A 54.5, pp. 3824–3851 (1996).
\bibitem{exp11}
Z.-H. Ma, Z.-H. Chen, J.-L. Chen, C. Spengler, A. Gabriel, and M. Huber, Measure of genuine multipartite entanglement with computable lower bounds, 
Phys. Rev. A {\bf 83}, 062325 (2011).









\bibitem{exp2}P. Hyllus, W. Laskowski, R.
Krischek, C. Schwemmer, W. Wieczorek, H. Weinfurter, L. Pezzé, and A. Smerzi, Fisher information and multiparticle entanglement. Phys. Rev. A {\bf 85}, 022321,
2012.
\bibitem{exp3}Z. Qin, M. Gessner, Z.
Ren, X. Deng, D. Han, W. Li, X. Su, A. Smerzi, and
K. Peng, Characterizing the multipartite
continuous-variable entanglement structure from
squeezing coefficients and the fisher information.
NPJ Quant Inf. 5, 3 (2019).
 
\bibitem{exp4}
M. Huber, F. Mintert, A. Gabriel, and B. C. Hiesmayr, Detection of High-Dimensional Genuine Multipartite Entanglement of Mixed States,
Phys. Rev. Lett. {\bf 104}, 210501 (2010).  

\bibitem{exp5}B. Lücke, J. Peise, G. Vitagliano, J. Arlt, L. Santos, G. Tóth, and C. Klempt. Detecting multiparticle entanglement of Dicke states. Phys. Rev. Lett. {\bf 112}, 155304 (2014).

\bibitem{exp6}V. Saggio, A. Dimić, C. Greganti, L. A. Rozema, P. Walther, and B. Dakić, Experimental few-copy multipartite entanglement detection,
Nature Phys. 15, 935–940 (2019).

\bibitem{exp7}H. Cao, S. Morelli, L. A. Rozema, C. Zhang, A. Tavakoli, and P. Walther, Genuine Multipartite Entanglement Detection with Imperfect Measurements: Concept and Experiment, Phys. Rev. Lett. {\bf 133}, 150201 (2024).

 \bibitem{exp8} Y.-Q. Zou, L.-N. Wu, Q. Liu, X.-Y. Luo, S.-F. Guo, J.-H. Cao, M. K. Tey, and L. You, Beating the classical precision limit with spin-1 Dicke states of more than 10,000 atoms, Proc.  Nat. Ac.  Sc. USA {\bf 115}, 6381–6385 (2018).

 
\bibitem{entmeasures6}
Measure of multipartite entanglement with computable lower bounds
Y. Hong, T. Gao, and F. Yan,
Phys. Rev. A 86, 062323 (2012).
\bibitem{entmeasures7}
 Y.-J. Luo, X. Leng, and C. Zhang,
 Genuine multipartite entanglement verification with convolutional neural networks,
Phys. Rev. A 110, 042412 (2024).
\bibitem{entmeasures8}
R. K. Malla, A. Weichselbaum, T.-C. Wei, and R. M. Konik,
Detecting Multipartite Entanglement Patterns Using Single-Particle Green’s Functions,
Phys. Rev. Lett.  133, 260202 (2024).
\bibitem{entmeasures9}F. Shi, L. Chen, G. Chiribella, and Q. Zhao, Entanglement Detection Length of Multipartite Quantum States,
Phys. Rev. Lett. 134, 050201 (2025). 
\bibitem{entmeasures10}S. Mukherjee, B. Mallick, S. Gopalkrishna Naik, A. G. Maity, and A. S. Majumdar, Detecting genuine multipartite entanglement using moments of positive maps,
Phys. Rev. A 112, 062428 (2025).

\bibitem{entmeasures11}P. Hyllus, W. Laskowski, R.
Krischek, C. Schwemmer, W. Wieczorek, H. Weinfurter, L. Pezzé, and A. Smerzi, Fisher information and multiparticle entanglement. Phys. Rev. A, 85, 022321,2012.



\bibitem{squash}M. Christandl and A.  Winter, ``Squashed entanglement'': an additive entanglement measure, J. Math.  Phys.  45.3, 829-840 (2004).


\bibitem{lieb}E. A. Carlen and E. H. Lieb, Bounds for Entanglement via an Extension of Strong Subadditivity of Entropy. Lett. Math. Phys. 101, 1–11 (2012).


\bibitem{hastings}M.M. Wolf, F. Verstraete, M. B. Hastings, and  J. I. Cirac, Area laws in quantum systems: mutual information and correlations, Phys. Rev. Lett. 100.7, 070502 (2008).

\bibitem{koashi_winter}M. Koashi and A.  Winter, Monogamy of quantum entanglement and other correlations, Phys. Rev. A 69.2, 022309 (2004).


\bibitem{quant_metro}V. Giovannetti, S. Lloyd, and L. Maccone, Quantum-enhanced measurements: beating the standard quantum limit, Science 306.5700, 1330-1336 (2004).

\bibitem{secret_sharing}M. Hillery, V. Bužek, and A. Berthiaume, Quantum secret
sharing, Phys. Rev. A 59.3, pp. 1829–1834 (1999).
\bibitem{W2}P. Agrawal and A. Pati, Perfect teleportation and superdense coding
with W states, Phys. Rev. A 74, 062320  (2006).

\bibitem{quant_telep}A. Karlsson and M. Bourennane, Quantum teleportation using
three-particle entanglement, Phys. Rev. A 58, pp. 4394–
4400 (1998).

\bibitem{exp1}O. Gühne and G. Tóth, Entanglement detection, Physics Reports
474.1-6, pp. 1–75 (2009).
\bibitem{exp0}B. M. Terhal, Detecting Quantum Entanglement, 	J. Th. Comp. Sci. 287(1), 313-335 (2002)

 
\bibitem{wootters}W. K. Wootters, Entanglement of formation of an arbitrary state of two qubits, Phys. Rev. Lett. {\bf 80}, 2245 (1998).

\bibitem{osborne}T. J. Osborne, Entanglement for rank-2 mixed states, Phys. Rev. A 72, 022309 (2005).

\bibitem{huang}H. Y. Huang, R. Kueng, and J. Preskill,  Predicting many properties of a quantum system from very few measurements. Nat. Phys. 16, 1050–1057 (2020).
\bibitem{science}T. Brydges, A. Elben, P. Jurcevic, B. Vermersch, C. Maier, B. P. Lanyon, P. Zoller, R. Blatt, and C. F. Roos	, Probing entanglement entropy via randomized measurements, Science 364, 260 (2019).
\bibitem{xiao}T. Zhang, G. Smith, J. A. Smolin, L. Liu, X.-J. Peng, Q. Zhao, D. Girolami, X. Ma, X. Yuan, and H. Lu, Quantification of Entanglement and Coherence with Purity Detection, NPJ Quantum Inf 10, 60 (2024).

\bibitem{daley}A. J. Daley, H. Pichler, J. Schachenmayer, and P. Zoller, Measuring entanglement growth in quench dynamics of bosons in an optical lattice,	Phys. Rev. Lett. 109, 020505 (2012).
 
\bibitem{henderson}L. Henderson and V. Vedral, Classical, quantum and total correlations, J. Phys. A 34, 6899 (2001).



\bibitem{eberly}T. Yu and J. H. Eberly, Evolution from Entanglement to Decoherence of Bipartite Mixed ``X'' States, Quant. Inf. and Comp. 7, 459-468 (2007).



\bibitem{dicke1}
M. Bergmann and O. Guehne, Entanglement criteria for Dicke states, 	J. Phys. A: Math. Theor. 46, 385304 (2013).
 
\bibitem{dicke2}J. K. Stockton, J. M. Geremia, A. C. Doherty, and H. Mabuchi, Characterizing the entanglement of symmetric many-particle spin-1/2 systems,
Phys.Rev. A, 69, 032109 (2004).




\bibitem{tufa}D. Girolami, T. Tufarelli, and C. E. Susa, Quantifying
Genuine Multipartite Correlations and their Pattern Complexity, Phys.
Rev. Lett. 119.14, 140505 (2017).





\end{thebibliography}
\end{document}